\documentclass{article}
\usepackage[T1]{fontenc} 
\usepackage[utf8]{inputenc} 
\usepackage{ismir,amsmath,cite,url}
\usepackage{graphicx}
\usepackage{color}
\usepackage{amssymb}
\usepackage{multirow}
\def\thickhline{\noalign{\hrule height.8pt}}

\usepackage{lineno}

\title{Combining audio control and style transfer using latent diffusion}

\multauthor
{Nils Demerlé \hspace{1cm} Philippe Esling \hspace{1cm} Guillaume Doras \hspace{1cm} David Genova}
  {Ircam, STMS Lab, Sorbonne Université, CNRS \\
{\tt\small demerle@ircam.fr, esling@ircam.fr, doras@ircam.fr, genova@ircam.fr}
}

\def\authorname{N. Demerlé, P. Esling, G. Doras, and D. Genova}

\begin{document}

\maketitle
\begin{abstract}

Deep generative models are now able to synthesize high-quality audio signals, shifting the critical aspect in their development from audio quality to control capabilities. Although text-to-music generation is getting largely adopted by the general public, explicit control and example-based style transfer are more adequate modalities to capture the intents of artists and musicians. 

In this paper, we aim to unify explicit control and style transfer within a single model by separating local and global information to capture musical structure and timbre respectively. To do so, we leverage the capabilities of diffusion autoencoders to extract semantic features, in order to build two representation spaces. We enforce disentanglement between those spaces using an adversarial criterion and a two-stage training strategy. Our resulting model can generate audio matching a timbre target, while specifying structure either with explicit controls or through another audio example. We evaluate our model on one-shot timbre transfer and MIDI-to-audio tasks on instrumental recordings and show that we outperform existing baselines in terms of audio quality and target fidelity. Furthermore, we show that our method can generate cover versions of complete musical pieces by transferring rhythmic and melodic content to the style of a target audio in a different genre. 

\end{abstract}
\section{Introduction}\label{sec:introduction}


Deep generative models are now particularly successful at synthesising high-quality, realistic audio signals. Hence, the major impediment to their broader use in creative workflows is not their audio quality anymore, but rather how end-users can have complete control over the generation process. Following early works on unconditional generation \cite{mehri2016samplernn, dhariwal2020jukebox}, multiple methods proposed to enable control by conditioning generation on semantic tags or audio-descriptors \cite{engel2020ddsp, devis2023continuous}. However, such supervised approaches remain limited to the use of explicit descriptors and are constrained by their reliance on annotated datasets. The recent development of language models and representation learning led to impressive performance in text-conditioned generation, mainly relying on transformers \cite{copet2024simple, agostinelli2023musiclm} or diffusion models \cite{huang2023noise2music, schneider2023mo, evans2024fast}. However, concepts such as timbre, musical style or genres boundaries are usually elusive and highly subjective. Hence, text descriptions might remain limited to common sounds and insufficient to precisely capture musical intentions. 

\begin{figure}[t]
 \centerline{
 \includegraphics[width=200pt]{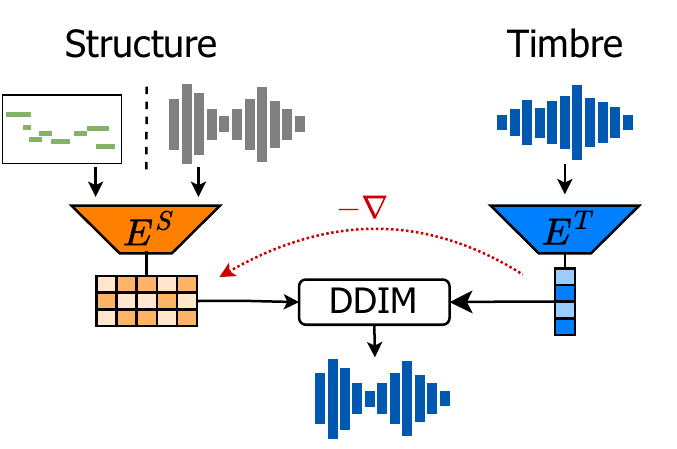}}
 \caption{General overview of our method. We extract timbre and structure representations from waveform and/or MIDI inputs using encoders $E_T$ ad $E_S$ respectively. Those representations condition a latent diffusion model, enabling both explicit and example-based control.}
 \label{fig:intro}
\end{figure}

A parallel stream of research to alleviate those issues is to guide specific aspects of the generation process by providing audio examples. Most approaches in this audio-to-audio editing are focused on \textit{timbre transfer}, where the timbre of a given sound is applied on the content of another. While some works can transfer any audio to the timbre of a given training set \cite{caillon2021rave}; others achieve many-to-many \textit{timbre transfer} but only between a small set of predefined instrument classes \cite{mor2018universal, bitton2020vector}. \textit{One-shot timbre transfer} between different instrument recordings have been achieved using Variational Autoencoders (VAE) \cite{luo2022towards, cifka2021self}, but these models rely on a latent bottleneck to enforce disentanglement between timbre and pitch, which hampers their ability to generate high-quality audio on real-world data. More recently, a text-inversion technique was proposed to perform musical style transfer between arbitrary content and style examples \cite{li2024music}, but it relies on a large pretrained text-to-music model and requires optimisation prior to each transfer, resulting in very slow inference. 

\newpage

In this paper, we aim to unify explicit control through audio descriptors or MIDI sequences and style transfer within a single model. To do so, we separate local, time-varying factors of variations and global information, capturing musical structure and timbre in two separate representation spaces. We slightly abuse the terms structure and timbre: by structure, we designate time-varying features, e.g. melody, loudness; by timbre, we designate global features such as actual timbre, but also style or genre. The principle of our method is depicted in Figure~\ref{fig:intro}.
Our approach is based on the recently proposed diffusion autoencoder \cite{preechakul2022diffusion}, which trains a semantic encoder to condition a diffusion model, in order to achieve both high-quality generation while being able to extract and control high-level features from the data. We extend this approach by building separate representations for timbre and structure, while enforcing their disentanglement with an adversarial criterion combined with a two-stage training strategy. Our method can generate audio matching a timbre target, while specifying the musical structure either with explicit controls (such as MIDI data input) or an audio example. For computation efficiency, our diffusion model operates in the latent space of pretrained autoencoders, resulting in faster than real-time inference on GPU\footnote{Experiments were conducted on a NVIDIA A5000 GPU}. 

First, we benchmark our model on a \textit{one-shot timbre transfer} tasks and demonstrate that our model improves upon existing baselines in terms of audio quality, timbre similarity as well as note onsets and pitch accuracy. On the same dataset, we show that our model can also generate audio from MIDI input and a target timbre example with performances superior to a state-of-the-art MIDI-to-audio baseline. 
Finally, we show that our method can be applied to complete musical pieces and generate cover versions of a track by transferring its rhythmic and melodic content to the style of a target audio in a different genre. We provide audio examples, additional experiments and source code on a supporting webpage\footnote{ \url{https://nilsdem.github.io/control-transfer-diffusion/}}

\section{Background}\label{sec:background}

\begin{figure*}[t]
 \centerline{
 \includegraphics[width=2.1\columnwidth]{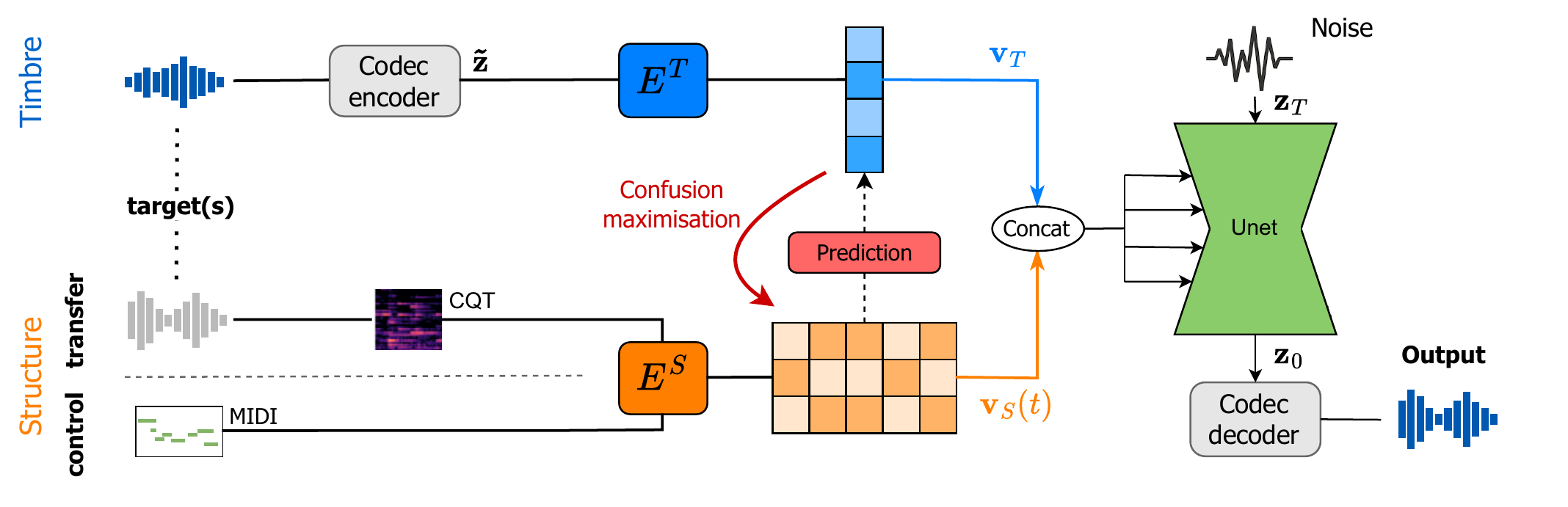}}
 \caption{Detailed overview of our method. Input signal(s) are passed to structure and timbre encoders, which provides semantic encodings that are further disentangled through confusion maximization. These are used to condition a latent diffusion model to generate the output signal. Input signals are identical during training and but distinct at inference.}
 \label{fig:main_method}
\end{figure*}

\subsection{Diffusion models}

Diffusion models (DMs) are a family of generative models that learn to reverse a stochastic process that gradually adds noise to the input data. These models benefit from high-quality generation, stable training and conditioning abilities, which led to their widespread adoption in image \cite{karras2022elucidating} and audio generation \cite{liu2023audioldm}. 

Formally, we define a \textit{forward process} $q(\mathbf{x}_{1:T}|\mathbf{x}_0) = \prod_{t=1}^T q(\mathbf{x}_t|\mathbf{x}_{t-1})$, which is a Markov chain that increasingly adds noise to the data $\mathbf{x}_0$ by relying on the conditional distribution 

\begin{equation}
q(\mathbf{x}_t | \mathbf{x}_{t-1}) = \mathcal{N}(\mathbf{x}_t; \sqrt{1 - \beta_t}\mathbf{x}_{t-1}, \beta_t \mathbf{I}),
\end{equation}

where $\beta_t$ are hyperparameters defining the noise levels at times $t \in \{0,T\}$, with $T \in \mathbb{N}$. We are interested in learning the reverse diffusion process $p_{\theta}(\mathbf{x}_{t-1} | \mathbf{x}_{t})$, from which we can iteratively denoise a random sample $\mathbf{x}_T \sim \mathcal{N}(0, \mathbf{I})$ to a data sample $\mathbf{x}_0 \sim p(\mathbf{x}_0)$. In a recenty study, \cite{ho2020denoising} made a connection between diffusion and denoising score matching \cite{song2019generative}, leading to a simplified formulation and improved experimental results. The authors show that the reverse process can be approximated by learning a denoising network $\epsilon_\theta$ that predicts the noise $\mathbf{\epsilon} \sim \mathcal{N}(\mathbf{\epsilon}, \mathbf{0}, \mathbf{I})$ used to corrupt the data. This results in a simpler training objective

\begin{equation}
    \min_{\theta \in \Theta} \quad \mathbb{E}_{t,\mathbf{x}_0,\epsilon} \big [ \| \epsilon_{\theta}(\sqrt{\alpha_t} \mathbf{x}_0 + \sqrt{1 - \alpha_t}\mathbf{\epsilon}, t) - \mathbf{\epsilon} \| \big ] ,
\end{equation}

where $\alpha_t = \prod_{s=1}^t (1 - \beta_s)$, and $\epsilon_{\theta}$ is usually parametrized by a UNet \cite{ronneberger2015u}. 

\subsection{Diffusion autoencoders}

DMs naturally yield a series of latent variables $\mathbf{x}_{1:T}$ through their \textit{forward process}. However, these stochastic variables built from increasingly adding noise do not capture much semantic information over the data. Although the more recent proposal of Denoising Diffusion Implicit Models (DDIMs) \cite{song2020denoising} extends the original diffusion formulation to a deterministic process allowing each data input to be mapped to a unique latent code $\mathbf{x}_T$, it still fails to extract and organise high-level features from of the data. Diffusion autoencoders \cite{preechakul2022diffusion} alleviate this issue by employing a learnable encoder that compress the data to a semantic latent code $\mathbf{z}_{sem} = E_{\phi} (\mathbf{x}_0)$, which then conditions a diffusion decoder. The semantic encoder and the DDIM decoder are trained jointly, following the objective

\begin{equation}
    \min_{\theta, \phi} \texttt{ } \mathbb{E}_{t,\mathbf{x}_0,\epsilon} \big [ \| \epsilon_{\theta}(\sqrt{\alpha_t} \mathbf{x}_0 + \sqrt{1 - \alpha_t}\epsilon, E_{\phi} (\mathbf{x}_0), t) - \epsilon \| \big ]
\end{equation}

On image applications, the authors show that the semantic code captures high-level attributes such as person identity, smile or presence of glasses, and can be used for downstream tasks such as conditional generation and attributes manipulation, while achieving state-of-art reconstruction. This approach was also successfully applied to audio \cite{schneider2023mo}, where the authors encode magnitude spectrograms into a semantic latent space, allowing to achieve high quality text-conditioned waveform generation.

\subsection{Control in audio generation}\label{sec:control}

A straightforward approach to extend unconditional generative models in order to provide instruments befitting artistic control is to introduce conditioning on explicit controls. The DDSP model \cite{engel2020ddsp} proposes explicit pitch and loudness conditioning, while FaderRave \cite{devis2023continuous} extended explicit control to non-differentiable time-varying attributes, but both methods remain limited to explicit descriptors and annotated datasets.  
While recent text-to-music methods like MusicGen \cite{copet2024simple} and Music ControlNet \cite{wu2023music}  have incorporated melody conditioning capabilities, their expressiveness remains constrained by the need to define subjective timbre properties through text prompts. Li et al. \cite{li2024music} proposed to use text-inversion in order to extract pseudo-words that represent timbre directly from audio, but their method is computationally intensive and requires to perform an optimisation for each new timbre target. Timbre conditioning directly from a waveform example was also recently proposed the context of bass accompaniment generation\cite{pasini2024bass}.



\subsection{Unsupervised disentanglement in sequential data}\label{sec:soa_disentanglement}

Many works proposed to model sequential data as a combination of local (time-variant) and global (time-invariant) factors of variation. Notably, the disentangled sequential autoencoder \cite{li2018disentangled} relies on simple architecture biases and parameter tuning to obtain disantangled local and global latent variables. Following this work, multiple methods improved the learned representation by explicitely minimizing the mutual information between the two learned variables \cite{bai2021contrastively, zhu2020s3vae, han2021disentangled}. It was shown that disentanglement can be further improved with contrastive learning and domain-specific transformations that preserve local or global attributes \cite{haga2023sequential, bai2021contrastively}.







More specifically in audio generation, SS-VAE \cite{cifka2021self} employs a Vector-Quantized VAE to achieve disentanglement through compression on quantized structure latent codes, combined with timbre-preserving data augmentations. Luo et al. \cite{luo2022towards} follow a two-stage training strategy similar to ours, and improve disentanglement by enforcing the consistency of the global and local latent variables in style or content transfer. However, both disentanglement strategies degrades reconstruction accuracy, which on top of spectrogram inversion based synthesis leads in poor audio quality. 

\section{Method}

Our approach is based on the assumption that musical audio samples can be seen as specific instances of a set of latent features that are separated between global features that capture style, and time-varying features that capture the local evolution of the signal. Although diffusion models are capable of high-quality conditional generation, they are computationally expensive when dealing with high-dimensional data. Hence, we employ an invertible audio codec to first compress the audio into a low-dimensional latent space, onto which we can build an efficient generative model. We extend the DiffAE \cite{preechakul2022diffusion} architecture to two semantic encoders in order to extract separately timbre and structure features from input samples. To further disentangle the learned features and improve transfer as well as explicit control performances, we employ an adversarial training strategy during training. In this section, we detail our proposed model depicted in Figure~\ref{fig:main_method}.

\subsection{Audio codec}

We build our audio codec as a convolutional autoencoder based on the RAVE model \cite{caillon2021rave} architecture, featuring the adversarial discriminator recently proposed in \cite{kumar2024high}. The model compresses audio waveforms $\mathbf{x}$ into an invertible latent sequence $\mathbf{z} \in \mathbb{R}^{L \times D}$, where $D$ and $L$ are the embedding space and time dimensions respectively. On top of the reconstruction and adversarial training objectives of RAVE, we introduce a penalty on the latent codes $f(\mathbf{z}) = max(0, |\mathbf{z}| - 1)$ to enforce that most latent codes are distributed between $-1$ and $1$. 

\subsection{Model structure}

We extract a timbre representation  $\mathbf{v}_T \in \mathbb{R}^{D_T}$ from an audio target using an encoder $E_\phi^{T}$ applied to the latent sequence $\mathbf{z}$ obtained with our audio codec. For structure, we extract a temporal representation $\mathbf{v}_S \in \mathbb{R}^{L\times D_S}$ from either an audio input or an explicit control signal $\mathbf{c}$ (such as a MIDI sequence), using an encoder $E_\psi^{S}$. In the case of an audio input, it would be natural to also infer it from the latent sequence $\mathbf{z}$. However, we found experimentally that it is particularly difficult to extract fine structure information from the highly-compressed representation $\mathbf{z}$. Hence, we instead infer structure form the Constant Q Transform (CQT) \cite{brown1991calculation} of the target signal, which has been shown to be a well-suited representation for pitch extraction tasks \cite{bittner2017deep}. For explicit control, the sequence $\mathbf{c}$ is directly used as input for $E_\psi^{S}$.

To generate audio, we sample a noise vector $\mathbf{z}_T$ and decode it to a latent code $\mathbf{z}_0$ through latent diffusion conditioned on representations $\mathbf{v}_S$ and $\mathbf{v}_T$. As diffusion formulation, we leverage the recent improvements introduced in \cite{karras2022elucidating} and parameterise our denoiser network $D_{\theta}$ to predict the data $\mathbf{z}_0$ instead of $\epsilon$. $E_\psi^{S}$, $E_\phi^{T}$ and $D_\theta$ are trained end-to-end to minimise the following loss function

\begin{equation}
    L_{diff} = \mathbb{E}_{t,\mathbf{z}_0,\epsilon} \| D_{\theta}(\sqrt{\alpha_t} \mathbf{z}_0 + \sqrt{1 - \alpha_t}\epsilon, \mathbf{v}_S, \mathbf{v}_T, t) - \mathbf{z}_0 \|
\end{equation}

We parameterise $D_{\theta}$ as a 1D convolutional UNet with residual blocks and self-attention layers. The two encoders share a  similar architecture as the encoding half of the UNet, with the difference that the timbre encoder compresses the input temporally and applies average pooling on the time dimension of the last layer. We condition the UNet architecture on $\mathbf{v}_S$ through concatenation with each block inputs. For the timbre vector \textbf{$\mathbf{v}_T$} we use Adaptative Group Normalisation (AdaGN) \cite{dhariwal2021diffusion}.

\subsection{Style and content disentanglement}\label{sec:disentanglement}
Although splitting the semantic content between two vectors that are constrained on their dimensions already encourages disentanglement between timbre and structure information, there is no theoretical guarantee regarding their separation. Furthermore, the appropriate feature dimensions are highly dependent on the task and dataset at hand. Hence, to enforce disentanglement without constraints on the dimensions, we introduce a two-stage training combined with an adversarial strategy. First, we freeze the structure encoder and train the model to build an adequate timbre representation. To avoid $\mathbf{v}_T$ to encode all of the information required to reconstruct the target $\mathbf{z}$, we extract timbre from a different sample $\mathbf{\tilde{z}}$ coming from the same track, following the assumption that it shares the same timbre as $\mathbf{z}$ but with a different structure. 

In the next stage, we introduce a discriminator $D_\zeta$ that tries to predict $\mathbf{v}_T$ from $\mathbf{v}_S$ and is trained to minimise 

\begin{equation}
L_D = \mathbb{E_{\mathbf{v}_S, \mathbf{v}_T}} \big [ \| \mathbf{v}_T - D_\zeta(\mathbf{v}_S) \| \big ]. 
\end{equation}

We train the discriminator alternatively with the encoders and denoising network, which try to minimize the following objective

\begin{equation}
L_{total} =  L_{diff} - \gamma L_D, 
\end{equation}

where $\gamma$ is an hyperparameter balancing between the reconstruction objective and disentanglement. Indeed, increasing $L_{D}$ maximises the confusion of the timbre information in the structure space. This restrains the diffusion model from reconstructing $\mathbf{z}$ solely from $\mathbf{v}_S$ and enables independant control of structure and timbre at inference. 

\section{Experiments}

We aim to asses the ability of our model to generate high-quality audio samples that match characteristics of structure and timbre targets, with the structure being either taken from an audio example or through an explicit control signal. 

\textbf{MIDI-to-audio} For explicit structure control, we evaluate the capability of our model to generate audio from a MIDI score and a target recording for timbre. We compare it to a state-of-art baseline in MIDI-to-audio generation. 

\textbf{Timbre transfer} We evaluate the efficiency of our disentanglement strategy on a task of \textit{one-shot timbre transfer} between polyphonic mono-instrument recordings. In this case, we consider that structure designates the notes being played (in terms of onset timing, pitch and loudness), while timbre corresponds to the remaining characteristics of the sound. We evaluate our model by randomly sampling timbre and structure examples and evaluate audio quality, timbre similarity as well as note accuracy. We compare our model with two example-based timbre transfer methods on synthetic and real recordings.

\subsection{Dataset}

\textbf{Synthetic Data} The Synthesized Lakh Dataset (SLAKH) \cite{manilow2019cutting} was generated from the LAKH MIDI collection using professional-grade sample-based virtual instruments. Synthesis parameters as well as audio effects settings were randomly chosen resulting in a very diverse set of timbres. We retain only the individual stems of non-percussive instruments, resulting in 400 hours of audio.

\noindent
\textbf{Real Data} To the best of our knowledge there is no multi-instrumental dataset of real recordings that contain a very large number of hours of audio. Hence, we combined the following three datasets to conduct our experiments :
\begin{itemize}
    \item \textbf{MaestroV2} : Maestro \cite{hawthorne2018enabling} is a piano dataset recorded on Disklavier pianos, capturing both audio and notes played, resulting in approximately 200hours of annotated piano recordings. 

     \item \textbf{GuitarSet} : Guitarset \cite{xi2018guitarset} is a collection of live guitar performances with solos and accompaniment from various genres and play styles, with a total audio duration of 6 hours.

     \item \textbf{URMP} : The URMP dataset \cite{li2018creating} is composed of pieces played by a large variety of classical instruments. For each piece we retain the mono-instrumental recordings, resulting in approximately 4 hours of audio.
\end{itemize}

As the GuitarSet and URMP are low sample-size datasets, we add synthetic data stems from SLAKH to the training set to facilitate learning. Furthermore, as the different datasets are greatly imbalanced in terms of sample size, we apply a sampling strategy to even the model performance on each dataset. Following \cite{hawthorne2022multi}, if $n_i$ is the number of samples in a given dataset, we draw examples from this dataset during training with probability $(n_i / \Sigma_j n_j)^{0.3}$.

\begin{table*}
\begin{tabular}{clccccccc}
\thickhline

\multicolumn{1}{l}{} &  & \multicolumn{2}{c}{\textbf{Quality (FAD)} $\downarrow$} & \multicolumn{2}{c}{\textbf{Timbre similarity} $\uparrow$} & \multicolumn{2}{c}{\textbf{Onset F1 score}$\uparrow$} \\ 
\multicolumn{1}{l}{} &  & \multicolumn{1}{c}{Rec.} & \multicolumn{1}{c}{Transfer} & \multicolumn{1}{c}{Rec.} & \multicolumn{1}{c}{Transfer} & \multicolumn{1}{c}{Rec.} & \multicolumn{1}{c}{Transfer} & \multicolumn{1}{c}{} \\ \hline
\multirow{3}{*}{\textbf{MIDI-to-audio}} & Spectrogram diffusion \cite{hawthorne2022multi} & 3.46 & - & 0.76 & - & 0.32 & - &\\
 & Ours w/o $E_S$ & 1.22 & 1.41 & 0.87 & 0.77 & \textbf{0.40} &  \textbf{0.38} \\
 & \textbf{Ours} & \textbf{0.88} & \textbf{1.06} & \textbf{0.89} & \textbf{0.83} & 0.36 & 0.23 \\ \hline
\multirow{6}{*}{\textbf{Timbre transfer}} &   SS-VAE\cite{cifka2021self} & 2.83 & 3.23 & 0.75 & 0.69 & 0.29 & 0.15 \\
 &  Music Style Transfer \cite{li2024music} & 2.95 & 2.77 & 0.84 & 0.60 & 0.22 & 0.17 &  \\
 & \text{Ours w/o adversarial - $D_S = 4$} & 0.95 & 1.75 & 0.91  & 0.75 & 0.36 & 0.23 \\
 & \text{Ours w/o adversarial - $D_S = 8$} & \textbf{0.95} &  1.65 & 0.91 & 0.73 & 0.36 & \textbf{0.26} \\
 & \textbf{Ours}  & 1.13 & \textbf{1.42} &  \textbf{0.91} & \textbf{0.82} & \textbf{0.36}  & 0.23 \\ \thickhline
\end{tabular}
\caption{Experimental results in terms of audio quality, timbre similarity and note accuracy on the SLAKH dataset, for MIDI-to-audio generation (\textbf{up}) and timbre transfer (\textbf{down}). "Rec." corresponds to samples generated from identical structure and timbre targets, while "Transfer" designates randomly chosen timbre targets.}
\label{table:synthetic}
\end{table*}

\subsection{Evaluation metrics}\label{sec:metrics}

We aim to evaluate how our method is able to match the timbre and structure targets characteristics, while maintaining high-quality audio. 

\textbf{Audio quality} We rely on the widely used \textit{Frechet Audio Distance} (FAD) \cite{kilgour2018fr} to evaluate how the generated audio distribution matches the dataset distribution, both for reconstructed and transferred samples. We use the available reference implementation of FAD\footnote{https://github.com/gudgud96/frechet-audio-distance/tree/main} and use VGGish \cite{hershey2017cnn} embeddings of the samples to compute the distance

\textbf{Timbre Similarity} To evaluate timbre similarity we employ the metric proposed in the SS-VAE baseline \cite{cifka2021self}. It relies on a triplet network trained to predict if samples are played by the same instrument based on Mel-frequency cepstral coefficients 2-13. We use their implementation and train the metric on the \textit{mixing-secrets}\footnote{https://www.cambridge-mt.com/ms/mtk/} dataset.

\textbf{Structure} To evaluate if our model is able to reproduce the notes of the structure target, we employ a transcription model \cite{bittner2022lightweight} and compare its output to the ground-truth MIDI data. As metric, we use Onset F1 score from \textit{mir-eval}, where two notes are considered identical if they have identical pitch and onsets within $\pm 50 ms$ of each other. 
\subsection{Baselines}

For the timbre transfer experiments, we compare our method to SS-VAE \cite{cifka2021self} and Music Style Transfer \cite{li2024music} presented in Section~\ref{sec:background}. We train both models on the real and synthetic datasets, using the official implementation. For explicit control, we evaluate our method against a MIDI-to-audio model \cite{hawthorne2022multi} that was also trained on the SLAKH dataset. 
We use the \textit{small} configuration of the publicly available pretrained model, as larger models do not fit on our NVIDIA A5000 GPU. 

\subsection{Training details}

We start by training our audio codec for 1M steps before training our diffusion model for 500k steps, with an initial timbre learning stage of 100k steps. The overall training takes one day on NVIDIA A5000 GPU. For all experiments we rely on the AdamW optimizer \cite{loshchilov2017decoupled} with a constant learning rate of $1e^-4$ and a batch size of 48. For inference we use the deterministic sampler proposed in \cite{karras2022elucidating} with 40 diffusion steps. 

\section{Results}

\subsection{MIDI-to-audio}

First, we evaluate our model performance in MIDI-to-audio generation in terms of audio quality, timbre similarity and Onset F1 score. We detail our results in Table~\ref{table:synthetic} for reconstruction and transfer setups, where the target timbre corresponds to either the instrument of the MIDI sequence or a different sample. In both cases, we obtain higher similarity, as our dedicated timbre embedding captures timbre much more precisely than simple label conditioning on instrument categories. Interestingly, we also obtain better F1 scores, although we did not design our model specifically for MIDI inputs as opposed to \cite{hawthorne2022multi} where authors employ a dedicated note sequence embedding strategy. 

To assess the benefit of our disentanglement strategy, we experiment with bypassing the structure encoder and directly conditioning the UNet on the MIDI sequence (\textit{Ours w/o $E_S$} entry in  Table~\ref{table:synthetic}). This results in overall better Onset F1 score, but degrades timbre similarity and FAD. This demonstrates that our disentanglement strategy improves the capability of the model to precisely render the timbre of the target recording. As described in Section~\ref{sec:metrics}, the Onset F1 score characterises the difference between the generated notes and the input MIDI sequence. The lower accuracy obtained with our full model in the transfer setup can be explained by the fact that some MIDI sequence are not plausible scores for some target instruments (such as playing chords with a flute). Through the disentangled structure encoding, our model is capable of adapting the input MIDI sequence to the range and capabilities of the target instrument, which results in more realistic sounding samples. We encourage the reader to listen to the examples on our supporting webpage that support this statement. 

\subsection{Timbre transfer}

\textbf{Synthetic data} Here, we first evaluate \textit{timbre transfer} on synthetic data and display the results in Table~\ref{table:synthetic}. The two baselines appear to provide low audio quality and timbre similarity, and both methods obtain lower Onset F1 scores indicating that they are not able to adequately control structure and timbre independently. Our method improve upon the baselines on all three evaluated aspects, and is interestingly able to reach a comparable performance as in the explicitly conditioned MIDI-to-audio setup.

We also performed an ablation study, by evaluating the effect of applying an information bottleneck on the structure latent space instead of using our adversarial strategy. As mentioned in Section~\ref{sec:disentanglement}, the model is able to transfer timbre when $D_S$ is small but achieves a low F1 score. Increasing the latent dimensions improves structure fidelity at the cost of degrading timbre similarity. Using our disentanglement strategy, we are able to employ a 32-dimensional latent space and achieve higher timbre similarity with a slight decrease in note accuracy. Although multiple definitions of timbre transfer are possible, we argue that the most convincing timbre transfer do not necessarily imply a perfect note structure F1 score. In the case of a transfer between monophonic instruments playing in the same pitch range, we can expect all the notes from one recording to be transferred to the other. However, when performing transfer between recordings with very distinct timbre such as a piano playing in its high range and a bass, an interesting transfer would rather be the bass playing the main melodic line a few octaves lower than the piano, which would result in a low F1 score. The improvement in terms of FAD between the distribution of transferred samples and the original data obtained with our disentanglement strategy supports that our method generates more realistic transfers, as an instrument playing notes outside its usual range would be considered as out-of-distribution.

\begin{table}[ht!]
\begin{tabular}{lccc}
\thickhline
\multicolumn{1}{l}{} &  \multicolumn{1}{c}{\textbf{FAD} $\downarrow$} & \multicolumn{1}{c}{\textbf{Timbre} $\uparrow$} & \multicolumn{1}{c}{\textbf{F1}$\uparrow$} \\

 SS-VAE\cite{cifka2021self} & 9.26 & 0.58 & 0.19   \\
Music Style Tr. \cite{li2024music} & 10.2 & 0.57 &  0.17 \\
 \text{Ours w/o adv.} & 2.14 &  0.81 & \textbf{0.43} \\
\textbf{Ours}  & \textbf{1.36} & \textbf{0.88} & 0.28   \\ 
\thickhline
\end{tabular}
\label{table:transfer_real}
\caption{Experimental results for timbre transfer on real instruments in terms of FAD, timbre similarity and onset F1 score. Metrics are averaged between the three datasets.}
\end{table}

\textbf{Real data}. We present our results for transfer between real instrumental recordings in Table~\ref{table:transfer_real}. Our model improves even further on the existing baselines for which real instruments timbre seems particularly challenging. Even without adversarial regularisation, our model obtains better FAD, timbre similarity and note accuracy. Our disentanglement strategy further improves timbre match, although the relative decrease in note accuracy appears to be greater than on synthetic data. We believe this is mainly due to a necessary simplification of note structure when transferring complex piano recordings to the mainly monophonic URMP instruments. The improvement of transfer quality captured by the FAD and timbre similarity supports this interpretation. 

Qualitatively, we found that on top of generating realistic samples with the appropriate structure, the model is also able to add characteristic sound artefacts of the target instrument such as fret or hammer noises, as well as matching precise acoustic features of the original recording such as reverb or background noise. 

\section{Complete Music Style Transfer}\label{sec:applications}

\begin{table}[ht!]
\begin{tabular}{lccc}
\thickhline
\multicolumn{1}{l}{} &  \multicolumn{1}{c}{\textbf{FAD $\downarrow$ }} & \multicolumn{1}{c}{\textbf{Cover (\%) $\uparrow$ }} & \multicolumn{1}{c}{\textbf{Genre $\uparrow$ }} \\
MusicGen \cite{copet2024simple} & - & 37.6 &  0.48  \\
\text{Ours w/o adv.} & 3.99 &  48.5 & 0.44 \\
\textbf{Ours}  & \textbf{3.31} & \textbf{52.2} & \textbf{0.55}  \\ 
\thickhline
\end{tabular}
\label{table:music_transfer}
\caption{Style transfer results on musical pieces, evaluated through FAD, cover identification and genre classification.}
\end{table}

Finally, we apply our model to the task of creating cover versions of a song that match the style of an example in a different genre. We rely on an in-house dataset of 200 hours of jazz, dub, lofi hip-hop and rock. We use the same model architecture to extract structure from the original track and timbre from the cover targets, with two minor modifications. First, we introduce temporal compression in the structure encoder to avoid $\mathbf{v}_T$ to capture information that is too precisely located in time. Second, we condition the UNet on a BPM time series to help it generating coherent rhythms. Without those modifications, the rhythmic elements from timbre and structure targets are conflicting with each other, resulting in somewhat chaotic generations. We evaluate our model using the cover detection algorithm proposed in \cite{abrassart2022and}, which outputs a cover probability based on melodic and harmonic similarities between tracks. To assess style transfer, we rely on the text and audio joint-embedding model CLAP \cite{wu2023large} and classify genre based on the cosine similarity between the audio and genre label embeddings. 
We compare our method to MusicGen \cite{copet2024simple}, a text-to-music generation model with audio-based melody conditioning. We derive an input prompt from the target genre and use the structure target for melody. 

Our model without regularisation obtains a better cover identification than MusicGen, and our disentanglement strategy further improves transfer resulting in higher genre accuracy. Qualitatively, MusicGen seems to only extract the main melodic idea from the structure audio, whereas our method is able to capture most of the harmonic and melodic content. Furthermore, as our model extract style directly form audio rather than through a text prompt, it transfers the different structural elements towards the instruments actually present in the timbre target rather than just the typical instruments of the genre.


\section{Conclusion}

We presented a simple method to learn disentangled timbre and structure representations. To the best of our knowledge, this is the first model capable of generating realistic, high-quality audio through transfer and MIDI rendering. We leave for future works improvements on the trade-off between reconstruction and disentanglement and applications to more complex musical datasets. Furthermore, we aim to initiate a reflection on the characterisation of the elusive concept of musical style transfer, which we believe to be an exciting stream of research towards a broader use of deep generative models in artistic work-flows. 

\bibliography{ISMIRtemplate}

\begin{thebibliography}{10}
\providecommand{\url}[1]{#1}
\csname url@samestyle\endcsname
\providecommand{\newblock}{\relax}
\providecommand{\bibinfo}[2]{#2}
\providecommand{\BIBentrySTDinterwordspacing}{\spaceskip=0pt\relax}
\providecommand{\BIBentryALTinterwordstretchfactor}{4}
\providecommand{\BIBentryALTinterwordspacing}{\spaceskip=\fontdimen2\font plus
\BIBentryALTinterwordstretchfactor\fontdimen3\font minus \fontdimen4\font\relax}
\providecommand{\BIBforeignlanguage}[2]{{%
\expandafter\ifx\csname l@#1\endcsname\relax
\typeout{** WARNING: IEEEtran.bst: No hyphenation pattern has been}%
\typeout{** loaded for the language `#1'. Using the pattern for}%
\typeout{** the default language instead.}%
\else
\language=\csname l@#1\endcsname
\fi
#2}}
\providecommand{\BIBdecl}{\relax}
\BIBdecl

\bibitem{mehri2016samplernn}
S.~Mehri, K.~Kumar, I.~Gulrajani, R.~Kumar, S.~Jain, J.~Sotelo, A.~Courville, and Y.~Bengio, ``Samplernn: An unconditional end-to-end neural audio generation model,'' \emph{arXiv preprint arXiv:1612.07837}, 2016.

\bibitem{dhariwal2020jukebox}
P.~Dhariwal, H.~Jun, C.~Payne, J.~W. Kim, A.~Radford, and I.~Sutskever, ``Jukebox: A generative model for music,'' \emph{arXiv preprint arXiv:2005.00341}, 2020.

\bibitem{engel2020ddsp}
J.~Engel, L.~Hantrakul, C.~Gu, and A.~Roberts, ``Ddsp: Differentiable digital signal processing,'' \emph{arXiv preprint arXiv:2001.04643}, 2020.

\bibitem{devis2023continuous}
N.~Devis, N.~Demerl{\'e}, S.~Nabi, D.~Genova, and P.~Esling, ``Continuous descriptor-based control for deep audio synthesis,'' in \emph{ICASSP 2023-2023 IEEE International Conference on Acoustics, Speech and Signal Processing (ICASSP)}.\hskip 1em plus 0.5em minus 0.4em\relax IEEE, 2023, pp. 1--5.

\bibitem{copet2024simple}
J.~Copet, F.~Kreuk, I.~Gat, T.~Remez, D.~Kant, G.~Synnaeve, Y.~Adi, and A.~D{\'e}fossez, ``Simple and controllable music generation,'' \emph{Advances in Neural Information Processing Systems}, vol.~36, 2024.

\bibitem{agostinelli2023musiclm}
A.~Agostinelli, T.~I. Denk, Z.~Borsos, J.~Engel, M.~Verzetti, A.~Caillon, Q.~Huang, A.~Jansen, A.~Roberts, M.~Tagliasacchi \emph{et~al.}, ``Musiclm: Generating music from text,'' \emph{arXiv preprint arXiv:2301.11325}, 2023.

\bibitem{huang2023noise2music}
Q.~Huang, D.~S. Park, T.~Wang, T.~I. Denk, A.~Ly, N.~Chen, Z.~Zhang, Z.~Zhang, J.~Yu, C.~Frank \emph{et~al.}, ``Noise2music: Text-conditioned music generation with diffusion models,'' \emph{arXiv preprint arXiv:2302.03917}, 2023.

\bibitem{schneider2023mo}
F.~Schneider, O.~Kamal, Z.~Jin, and B.~Sch{\"o}lkopf, ``Mousai: Text-to-music generation with long-context latent diffusion,'' \emph{arXiv preprint arXiv:2301.11757}, 2023.

\bibitem{evans2024fast}
Z.~Evans, C.~Carr, J.~Taylor, S.~H. Hawley, and J.~Pons, ``Fast timing-conditioned latent audio diffusion,'' \emph{arXiv preprint arXiv:2402.04825}, 2024.

\bibitem{caillon2021rave}
A.~Caillon and P.~Esling, ``Rave: A variational autoencoder for fast and high-quality neural audio synthesis,'' \emph{arXiv preprint arXiv:2111.05011}, 2021.

\bibitem{mor2018universal}
N.~Mor, L.~Wolf, A.~Polyak, and Y.~Taigman, ``A universal music translation network,'' \emph{arXiv preprint arXiv:1805.07848}, 2018.

\bibitem{bitton2020vector}
A.~Bitton, P.~Esling, and T.~Harada, ``Vector-quantized timbre representation,'' \emph{arXiv preprint arXiv:2007.06349}, 2020.

\bibitem{luo2022towards}
Y.-J. Luo, S.~Ewert, and S.~Dixon, ``Towards robust unsupervised disentanglement of sequential data--a case study using music audio,'' \emph{arXiv preprint arXiv:2205.05871}, 2022.

\bibitem{cifka2021self}
O.~C{\'\i}fka, A.~Ozerov, U.~{\c{S}}im{\c{s}}ekli, and G.~Richard, ``Self-supervised vq-vae for one-shot music style transfer,'' in \emph{ICASSP 2021-2021 IEEE International Conference on Acoustics, Speech and Signal Processing (ICASSP)}.\hskip 1em plus 0.5em minus 0.4em\relax IEEE, 2021, pp. 96--100.

\bibitem{li2024music}
S.~Li, Y.~Zhang, F.~Tang, C.~Ma, W.~Dong, and C.~Xu, ``Music style transfer with time-varying inversion of diffusion models,'' in \emph{Proceedings of the AAAI Conference on Artificial Intelligence}, vol.~38, no.~1, 2024, pp. 547--555.

\bibitem{preechakul2022diffusion}
K.~Preechakul, N.~Chatthee, S.~Wizadwongsa, and S.~Suwajanakorn, ``Diffusion autoencoders: Toward a meaningful and decodable representation,'' in \emph{Proceedings of the IEEE/CVF Conference on Computer Vision and Pattern Recognition}, 2022, pp. 10\,619--10\,629.

\bibitem{karras2022elucidating}
T.~Karras, M.~Aittala, T.~Aila, and S.~Laine, ``Elucidating the design space of diffusion-based generative models,'' \emph{Advances in Neural Information Processing Systems}, vol.~35, pp. 26\,565--26\,577, 2022.

\bibitem{liu2023audioldm}
H.~Liu, Z.~Chen, Y.~Yuan, X.~Mei, X.~Liu, D.~Mandic, W.~Wang, and M.~D. Plumbley, ``Audioldm: Text-to-audio generation with latent diffusion models,'' \emph{arXiv preprint arXiv:2301.12503}, 2023.

\bibitem{ho2020denoising}
J.~Ho, A.~Jain, and P.~Abbeel, ``Denoising diffusion probabilistic models,'' \emph{Advances in neural information processing systems}, vol.~33, pp. 6840--6851, 2020.

\bibitem{song2019generative}
Y.~Song and S.~Ermon, ``Generative modeling by estimating gradients of the data distribution,'' \emph{Advances in neural information processing systems}, vol.~32, 2019.

\bibitem{ronneberger2015u}
O.~Ronneberger, P.~Fischer, and T.~Brox, ``U-net: Convolutional networks for biomedical image segmentation,'' in \emph{Medical image computing and computer-assisted intervention--MICCAI 2015: 18th international conference, Munich, Germany, October 5-9, 2015, proceedings, part III 18}.\hskip 1em plus 0.5em minus 0.4em\relax Springer, 2015, pp. 234--241.

\bibitem{song2020denoising}
J.~Song, C.~Meng, and S.~Ermon, ``Denoising diffusion implicit models,'' \emph{arXiv preprint arXiv:2010.02502}, 2020.

\bibitem{wu2023music}
S.-L. Wu, C.~Donahue, S.~Watanabe, and N.~J. Bryan, ``Music controlnet: Multiple time-varying controls for music generation,'' \emph{arXiv preprint arXiv:2311.07069}, 2023.

\bibitem{pasini2024bass}
M.~Pasini, M.~Grachten, and S.~Lattner, ``Bass accompaniment generation via latent diffusion,'' in \emph{ICASSP 2024-2024 IEEE International Conference on Acoustics, Speech and Signal Processing (ICASSP)}.\hskip 1em plus 0.5em minus 0.4em\relax IEEE, 2024, pp. 1166--1170.

\bibitem{li2018disentangled}
Y.~Li and S.~Mandt, ``Disentangled sequential autoencoder,'' \emph{arXiv preprint arXiv:1803.02991}, 2018.

\bibitem{bai2021contrastively}
J.~Bai, W.~Wang, and C.~P. Gomes, ``Contrastively disentangled sequential variational autoencoder,'' \emph{Advances in Neural Information Processing Systems}, vol.~34, pp. 10\,105--10\,118, 2021.

\bibitem{zhu2020s3vae}
Y.~Zhu, M.~R. Min, A.~Kadav, and H.~P. Graf, ``S3vae: Self-supervised sequential vae for representation disentanglement and data generation,'' in \emph{Proceedings of the IEEE/CVF Conference on Computer Vision and Pattern Recognition}, 2020, pp. 6538--6547.

\bibitem{han2021disentangled}
J.~Han, M.~R. Min, L.~Han, L.~E. Li, and X.~Zhang, ``Disentangled recurrent wasserstein autoencoder,'' \emph{arXiv preprint arXiv:2101.07496}, 2021.

\bibitem{haga2023sequential}
T.~Haga, H.~Kera, and K.~Kawamoto, ``Sequential variational autoencoder with adversarial classifier for video disentanglement,'' \emph{Sensors}, vol.~23, no.~5, p. 2515, 2023.

\bibitem{kumar2024high}
R.~Kumar, P.~Seetharaman, A.~Luebs, I.~Kumar, and K.~Kumar, ``High-fidelity audio compression with improved rvqgan,'' \emph{Advances in Neural Information Processing Systems}, vol.~36, 2024.

\bibitem{brown1991calculation}
J.~C. Brown, ``Calculation of a constant q spectral transform,'' \emph{The Journal of the Acoustical Society of America}, vol.~89, no.~1, pp. 425--434, 1991.

\bibitem{bittner2017deep}
R.~M. Bittner, B.~McFee, J.~Salamon, P.~Li, and J.~P. Bello, ``Deep salience representations for f0 estimation in polyphonic music.'' in \emph{ISMIR}, 2017, pp. 63--70.

\bibitem{dhariwal2021diffusion}
P.~Dhariwal and A.~Nichol, ``Diffusion models beat gans on image synthesis,'' \emph{Advances in neural information processing systems}, vol.~34, pp. 8780--8794, 2021.

\bibitem{manilow2019cutting}
E.~Manilow, G.~Wichern, P.~Seetharaman, and J.~Le~Roux, ``Cutting music source separation some slakh: A dataset to study the impact of training data quality and quantity,'' in \emph{2019 IEEE Workshop on Applications of Signal Processing to Audio and Acoustics (WASPAA)}.\hskip 1em plus 0.5em minus 0.4em\relax IEEE, 2019, pp. 45--49.

\bibitem{hawthorne2018enabling}
C.~Hawthorne, A.~Stasyuk, A.~Roberts, I.~Simon, C.-Z.~A. Huang, S.~Dieleman, E.~Elsen, J.~Engel, and D.~Eck, ``Enabling factorized piano music modeling and generation with the maestro dataset,'' \emph{arXiv preprint arXiv:1810.12247}, 2018.

\bibitem{xi2018guitarset}
Q.~Xi, R.~M. Bittner, J.~Pauwels, X.~Ye, and J.~P. Bello, ``Guitarset: A dataset for guitar transcription.'' in \emph{ISMIR}, 2018, pp. 453--460.

\bibitem{li2018creating}
B.~Li, X.~Liu, K.~Dinesh, Z.~Duan, and G.~Sharma, ``Creating a multitrack classical music performance dataset for multimodal music analysis: Challenges, insights, and applications,'' \emph{IEEE Transactions on Multimedia}, vol.~21, no.~2, pp. 522--535, 2018.

\bibitem{hawthorne2022multi}
C.~Hawthorne, I.~Simon, A.~Roberts, N.~Zeghidour, J.~Gardner, E.~Manilow, and J.~Engel, ``Multi-instrument music synthesis with spectrogram diffusion,'' \emph{arXiv preprint arXiv:2206.05408}, 2022.

\bibitem{kilgour2018fr}
K.~Kilgour, M.~Zuluaga, D.~Roblek, and M.~Sharifi, ``Fr$\backslash$'echet audio distance: A metric for evaluating music enhancement algorithms,'' \emph{arXiv preprint arXiv:1812.08466}, 2018.

\bibitem{hershey2017cnn}
S.~Hershey, S.~Chaudhuri, D.~P. Ellis, J.~F. Gemmeke, A.~Jansen, R.~C. Moore, M.~Plakal, D.~Platt, R.~A. Saurous, B.~Seybold \emph{et~al.}, ``Cnn architectures for large-scale audio classification,'' in \emph{2017 ieee international conference on acoustics, speech and signal processing (icassp)}.\hskip 1em plus 0.5em minus 0.4em\relax IEEE, 2017, pp. 131--135.

\bibitem{bittner2022lightweight}
R.~M. Bittner, J.~J. Bosch, D.~Rubinstein, G.~Meseguer-Brocal, and S.~Ewert, ``A lightweight instrument-agnostic model for polyphonic note transcription and multipitch estimation,'' in \emph{ICASSP 2022-2022 IEEE International Conference on Acoustics, Speech and Signal Processing (ICASSP)}.\hskip 1em plus 0.5em minus 0.4em\relax IEEE, 2022, pp. 781--785.

\bibitem{loshchilov2017decoupled}
I.~Loshchilov and F.~Hutter, ``Decoupled weight decay regularization,'' \emph{arXiv preprint arXiv:1711.05101}, 2017.

\bibitem{abrassart2022and}
M.~Abrassart and G.~Doras, ``And what if two musical versions don't share melody, harmony, rhythm, or lyrics?'' in \emph{ISMIR 2022}, 2022.

\bibitem{wu2023large}
Y.~Wu, K.~Chen, T.~Zhang, Y.~Hui, T.~Berg-Kirkpatrick, and S.~Dubnov, ``Large-scale contrastive language-audio pretraining with feature fusion and keyword-to-caption augmentation,'' in \emph{ICASSP 2023-2023 IEEE International Conference on Acoustics, Speech and Signal Processing (ICASSP)}.\hskip 1em plus 0.5em minus 0.4em\relax IEEE, 2023, pp. 1--5.

\end{thebibliography}

%
%
%
%
%

\end{document}